\documentclass[12pt]{article}
 \usepackage{amssymb,amsfonts,latexsym}
\def\be            {\begin{equation}}
\def\bc       {boundary condition}

\def\calc  {{\cal C}}
\def\cald  {{\cal D}}
\def\calh  {{\cal H}}
\def\cals  {{\cal S}}
\def\complex       {{\mathbb C}}

\def\Hom           {{\rm Hom}}
\newcommand\hsp[1] {\mbox{\hspace{#1 em}}}

\def\iN            {\,{\in}\,}
\def\M             {{\rm M}}
\newcommand\nxt[1] {\\\raisebox{.12em}{\rule{.35em}{.35em}}\hsp{.6}#1}
\def\reals         {{\mathbb R}}

\def\zet           {{\mathbb Z}}
\begin {document}

\begin{flushright}  {\sf HU-EP-03-11} \\[-1cm]
{\sf hep-th/0302200} \end{flushright}

\begin{center} \vskip 19mm
{\Large\bf BOUNDARIES, DEFECTS AND}\\[4mm]
{\Large\bf FROBENIUS ALGEBRAS}\\[27mm]
{\large J\"urgen Fuchs $^1$ \ Ingo Runkel $^{2,3}$ \
and \  Christoph Schweigert $^{2,4}$ \\[13mm]
{\small $^1$ Institutionen f\"or fysik, Universitetsgatan 5,
\,S\,--\,651\,88\, Karlstad \\[2mm]
$^2$
LPTHE, Universit\'e Paris VI, 4 place Jussieu,
\,F\,--\,75\,252\, Paris\, Cedex 05 \\[2mm]
$^3$
Institut f\"ur Physik, HU Berlin,
Invalidenstra\ss{}e 110, \,D\,--\,10115 Berlin  \\[2mm]
$^4$
Institut f\"ur theoretische Physik E, RWTH Aachen, Sommerfeldstr.\ 28,
\,D\,--\,52074 Aachen
}
}
\end{center}
\vskip 23mm
\begin{quote}{\bf Abstract}\\[1mm]
The interpretation of D-branes in terms of open strings has lead to much 
interest in boundary conditions of two-dimensional conformal field theories 
(CFTs). These studies have deepened our understanding of CFT and allowed
us to develop new computational tools. The construction of CFT correlators 
based on combining tools from topological field theory and non-commutative
algebra in tensor categories, which we summarize in this contribution,
allows e.g.\ to discuss, apart from boundary conditions, also 
defect lines and disorder fields.
\end{quote}

  \newpage %%\vskip11mm 

\section{Basic CFT facts}

In the applications of two-dimensional conformal field theory
-- two-dimensional critical systems, quasi one-dimensional condensed
matter physics, and string theory -- boundary effects play an important role. 
They appear in the description of pointlike defects such as the Kondo 
effect, the computation of percolation probabilities, the construction 
of type I superstring theories, and the formulation of open string
perturbation theory in the background of D-branes. The study of boundary
conditions has very much improved our understanding of the structure of 
CFT; this is one topic we address in the present contribution. But 
two-dimensional conformal field theories possess an extremely rich structure,
which is in fact not exhausted by bulk and boundary fields. In particular,
they also involve defect lines and disorder fields; these
objects will be discussed as well. Hitherto, they have not yet played a 
role in world sheet theories of strings.

\vskip2mm

Let us briefly review a few pertinent aspects of CFT.
In the bulk there are both left moving and right
moving degrees of freedom, which are described by a left and a right moving 
chiral algebra, respectively. Each of them contains the Virasoro algebra,
which realizes the conformal symmetry, given by
generators $\{L_n\,{\mid}\, n\iN\mathbb Z\}$ with relations
  $$ [L_n,L_m] = (n{-}m)\, L_{n+m} + \mbox{\Large$\frac c{12}$}\,n\,(n^2{-}1)
  \,\delta_{n,-m} \,. $$

It is natural to study the representations of the chiral algebra; they
provide the superselection sectors of the chiral CFT. Primary fields 
correspond to irreducible representations; if their number is finite, 
the theory is called {\em rational\/}. With the help of the representation 
theory one can introduce conformal blocks 
which are spaces of multivalued functions
of insertion points on a complex curve. For WZW theories, they can be 
computed as (subspaces of) the spaces of solutions to Knizhnik-Zamolodchikov
equations.  For the full local conformal field theory they play the role 
of {\em pre\/}correlators; they also form the state spaces of the associated
three-dimensional topological field theory. The monodromy properties of 
these conformal blocks provide us with a collection of data -- fusing 
matrices, braiding matrices, conformal weights, representations of 
mapping class groups -- that encode the basic information about the chiral 
CFT; they are usually referred to as chiral data, or also as Moore-Seiberg 
data. 

A good handle on these data is an essential prerequisite for any powerful 
approach to CFT. Technically, it is afforded by the so-called
{\em modular tensor categories\/}. They furnish a basis-independent
formalization of the chiral data, which are obtained from solutions to 
differential equations, in much the same way as the
notion of a vector space constitutes a ba\-sis-in\-dependent description
for the structure of the space of solutions to linear equations.

The representations of a chiral algebra form a {\em category\/}; its 
objects are representations, its morphisms are in\-ter\-twining maps between 
representations. Here the term `chiral algebra' should be taken with
some care -- we do not necessarily start with a maximally extended chiral
algebra, for left movers or for right movers. This gives us sufficient
flexibility to address symmetry breaking boundary conditions as
well as situations in which left moving and right moving chiral algebra
do not coincide. 

\noindent
The representation category of a chiral algebra comes with a lot of structure:
\nxt   The notion of a tensor product, or fusion, of representations
leads to a tensor product 
  $$ \otimes\,: \quad \calc\times \calc \to \calc $$
for the category. For this tensor product, the vacuum sector acts as a unit.
\nxt   Braid group statistics in low dimensions provides us with the notion 
of a braiding, which constitutes a systematic prescription to permute 
representations in tensor products that
is compatible with the braid group relations.
\nxt  The concept of dual representations gives rise to a pairing of
representations.
\\[.5mm]
These structures obey a number of compatibility relations, and these are
summarized in the definition of a modular tensor category. The
qualification `modular' indicates the 
fact that from such a category one can infer a natural candidate for the 
action of the modular group, i.e.\ the mapping class group of the torus,
on the characters of the CFT. Modular tensor categories also arise as 
(truncations of) categories of representations of quantum groups, and 
as categories of representations of weak Hopf algebras.
It is an important result \cite{TUra} that to
every modular tensor category one can associate a topological field
theory in three dimensions. This realizes a ``holographic'' correspondence 
between {\em chiral\/} CFT in two dimensions and TFT in three dimensions.

\section{Algebra and topological field theory}

The central idea of our approach to (rational) conformal field theory
is to combine tools from topological field theory with non-commutative
algebra in tensor categories. \\
The central insight can be summarized as follows:

\vskip2mm
\framebox{ \begin{tabular}l
For a given chiral algebra, a full local CFT can be 
constructed from the \\ algebra
of open string states for a single boundary condition (\,= D-brane).
\end{tabular} }
    
    \vskip6mm
\noindent
To understand this result, let us examine the algebra of open string 
states, i.e.\ the algebra of boundary operators $\Psi_a^{AA}$
that respect a given boundary condition $A$, in more detail: 
\nxt The field-state correspondence allows us to work with the 
corresponding states. They form a module over the chiral algebra, i.e.\ 
a (generically reducible) object in the modular tensor category $\calc$;
we use the symbol $A$ also to denote this object.
\nxt The operator product
  $$ \Psi_a^{AA}(x)\, \Psi_b^{AA}(y) \,\sim\,
  \sum_c {\sf C}_{a \, b}^c\, (x{-}y)^{\Delta_c-\Delta_a-
  \Delta_b}\, \Psi_c^{AA}(y) $$
of boundary fields is associative. This leads to an equation of the type
  $$  {\sf C} \, {\sf C} \,\,=\,\, \mbox{\large\raisebox{.15em}{$\sum$}}
  \,{\sf C}\, {\sf C}\, F\,. $$ 
It is a first crucial observation that this relation gives rise to an 
associative multiplication morphism $m\iN\Hom(A{\otimes} A,A)$ on the 
object $A$. The presence of the fusing matrix $F$ in the relation for the 
structure constants {\sf C} is taken into account by a non-trivial notion 
of associativity in the category $\calc$.
\nxt 
Another important aspect of the algebra object $A$ is the fact that the
non-degeneracy of the two-point functions of boundary fields on the disk 
implies that $A$ can be endowed with the structure of a Frobenius algebra, 
i.e.\ an algebra with a non-degenerate invariant bilinear form.
(For precise definitions, see \cite{fuRs4}). Actually, $A$ is
even a so-called {\em symmetric special Frobenius algebra\/}. It should be
appreciated, though, that $A$ is not necessarily (braided-)\,commutative, 
as befits an algebra of boundary fields. Moreover, the boundary
condition we start with need not be elementary; it may equally well be 
a superposition of several elementary boundary conditions, including 
possibly Chan-Paton multiplicities.
\nxt
Finally, for any other boundary condition $M$ the
associativity of the operator product
  $$ \Psi^{AA}\, \Psi^{AA}\, \Psi^{AM} $$
of boundary fields
allows us to abstract a map $\rho\iN\Hom(A{\otimes} M, M)$ that endows the
corresponding object $M$ of $\calc$ with the structure of a (left) 
$A$-{\em module\/}.  We thus learn that boundary conditions are given by 
$A$-modules.

\vskip2mm
Let us now discuss the second input of our construction, topological field
theory. As a direct consequence of the axioms, every modular tensor category 
allows for the construction of a three-dimensional TFT. 
The latter assigns to every closed oriented two-manifold 
$\hat \Sigma$ a finite-dimensional vector space
$\calh(\hat \Sigma)$, the space of conformal blocks, on which the mapping 
class group $M\!ap(\hat \Sigma)$ acts projectively. 
Further, to every three-manifold with boundary $\hat \Sigma$ that 
contains a Wilson graph, the TFT assigns a vector in $\calh(\hat \Sigma)$.
The axioms of a TFT formalize the well-known relation between a 
TFT on a three-dimensional manifold and a {\em chiral\/} CFT on its boundary,
a structure that is e.g.\ central to the CFT description of universality 
classes of quantum Hall fluids (for a review, see e.g.\ \cite{fpsw}).

\section{The TFT construction of CFT correlators}

Our goal is now to determine CFT correlators on a surface $\Sigma$ that 
possibly has a boundary. To be able to apply tools from TFT,
we need to work with a suitable two-manifold
$\hat\Sigma$ without boundary. A natural candidate for $\hat \Sigma$ is 
the complex double \cite{bcdcd} of $\Sigma$. It comes with an orientation 
reversing involution $\sigma$ such that $\Sigma$ is identified with the
quotient of $\hat \Sigma$ by the action of $\sigma$. 
Indeed, it is natural \cite{scfu4} to view the world sheet 
$\Sigma$ as a real scheme and $\hat \Sigma$ as its complexification; the 
orientifold map $\sigma$ is then just the action of the non-trivial
element of the Galois group $Gal(\complex/\reals)\,{\cong}\,\zet_2$.

With the help of the double cover $\hat \Sigma$, we can in particular 
give a concise version 
of the principle of {\em holomorphic factorization\/}: The correlators 
of the CFT on $\Sigma$ are specific vectors in the space $\calh(\hat\Sigma)$
of conformal blocks on the complex double, which are determined by
two types of constraints: \\[1mm]
(1) They must be invariant under the action of $M\!ap(\Sigma) \,{\cong}\, 
    M\!ap(\hat\Sigma{)}_{\phantom|}^\sigma$. \\[.7mm]
(2) They must satisfy factorization rules.

\vskip2mm
According to the principle of holomorphic factorization we need to select a 
specific element of $\calh(\hat \Sigma)$ to describe a correlator on $\Sigma$. 
By the 
principles of TFT, in turn, such a vector is determined by a three-manifold 
$\M_\Sigma$ whose boundary is the double $\hat\Sigma$ and a ribbon graph in 
$\M_\Sigma$. For $\M_\Sigma$ we take the so-called {\em connecting manifold\/}
\cite{fffs2}, defined as the quotient of the interval bundle 
$\hat \Sigma\,{\times}\,[-1{,}1]$ by the $\zet_2$ that acts as $\sigma$
on $\hat \Sigma$ and as $t\,{\mapsto}\, {-}t$ on the interval. The points 
with $t\,{=}\,0$ provide a distinguished embedding of $\Sigma$ into 
$\M_\Sigma$. In fact, $\Sigma$ is a retract of $\M_\Sigma$ or, in more 
intuitive terms, $\M_\Sigma$ is just a fattening 
of the world sheet $\Sigma$. (To give one example: the double of a disk is
a sphere, and the orientifold map is the reflection at the equatorial
plane; the connecting manifold is in this case a full three-ball.) Concerning
the prescription for the ribbon graph in $\M_\Sigma$ we refer to
\cite{fuRs}. We merely recall that it involves a (dual) triangulation 
of $\Sigma$ with the ribbons labelled by the Frobenius algebra object $A$
and with the trivalent vertices corresponding to the multiplication morphism
$m$ of $A$.

In this framework, the consistency relations of modular invariance (1)
and factorization (2) can be proven rigorously. We also recover the various
combinatorial data -- such as modular invariant partition functions, 
NIM-reps and classifying algebras -- that arise when studying specific
small subsets of these consistency conditions
(see e.g.\ \cite{prss3,fuSc5,zbpp}). Moreover, Morita equivalence, 
combined with orbifold technology, allows for an elegant proof of 
T-dualities for arbitrary topology of the world sheet.

We have already seen that modules over $A$ correspond to boundary conditions. 
In non-commutative algebra, it is natural to consider {\em bimodules\/} as 
well, and again they turn out to possess a physical interpretation: they
describe defect lines. More specifically, these defect lines carry neither 
momentum nor charges. Thus they are topological in nature, so that their 
position matters only up to homotopy. These defect lines can actually be 
generalized to interfaces between two different full local CFTs 
that are based on one and the same chiral CFT.
Disorder fields create or change defect lines. 

The following table, associating algebraic structures to physical concepts,
can serve as a succinct summary of our results \cite{fuRs4}:
\begin{center}
\begin{tabular}{l|ll}
\ \ Physical concept & \ \ \ \ Algebraic structure \ \  \\[3pt]
\cline{1-2}\\[-9pt]
Boundary condition & \,left $A$-module \\[1pt]
Boundary field $\Psi_a^{MN}$ & $\,\Hom_A(M{\otimes}a, N)$ \\[2pt]
\cline{1-2}\\[-9pt]
Bulk field $\Phi_{ab}$ & 
\multicolumn2{|l} {$\,\Hom_{A,A}( (A{\otimes}a)^-\!,(A{\otimes}b)^+) $} 
\\[1.5pt]
Defect line & $\,A$-bimodule \\[1pt]
Disorder field $\Phi_{ab}^{B_1B_2}$ & 
\multicolumn2{|l} {$\,\Hom_{A,A}( (B_1{\otimes}a)^-\!,(B_2{\otimes}b)^+) $} 
\end{tabular}
\end{center}
\smallskip
In particular, bulk fields can be regarded as a special type of disorder 
fields -- those which connect the trivial defect line, i.e.\ the one labelled 
by $A$, to the trivial defect line.

\section{An example: Permutation branes}

Our algebraization of physical concepts not only leads to statements 
that can be proved rigorously, but also allows us to establish
powerful algorithms for doing computations. In particular, for constructing
a full local CFT only a single non-linear constraint needs to be solved:
the one that encodes associativity of the Frobenius algebra object $A$. 
We illustrate the power of our approach with the following example.

Consider the tensor product of $N$ identical conformal field theories
whose modular tensor category is $\calc$. We are then dealing with
the category $\calc^{\times N}$, which comes with an action of the 
symmetric group $S_N$. %%  on $N$ letters.
 One may wish to classify all those boundary conditions of
the $\calc^{\times N}$-theory which respect the subalgebra of the chiral 
algebra that is left pointwise fixed under the $S_N$-action. 
To perform this task we can work with the orbifold category
  $$ \calc_{\rm orb} := \calc^{\times N} \!/ S_N $$
for which many aspects of the chiral data are explicitly known
\cite{bohs,bantx}. In particular, the primary fields of the orbifold theory
are labelled by pairs $(\{\vec\lambda\},\psi)$, where $\{\vec\lambda\}$ is 
an unordered $N$-tuple of labels for simple objects in $\calc$ and $\psi$ is 
a character of the double of the stabilizer subgroup 
$S_{\vec\lambda}$ of $\{\vec\lambda\}$ in $S_N$.

It is not difficult to find the relevant algebra object. The algebra 
structure is determined by the product in the algebra of functions on $S_N$, 
and it is realized on the object
  $$ A=\bigoplus_{\psi\in S_N^*}\dim V_{\psi}\,(\{\vec 0\},(\psi,[e]))\,. $$
Here $\vec 0$ is the $N$-tuple each entry of which is the vacuum sector of 
$\calc$. Its stabilizer is all of $S_N$; the pairs $(\psi,[e])$,
consisting of a character $\psi$ of $S_N$ and the conjugacy class of the unit 
element $e\iN S_N$, specify a subclass of
irreducible representations of the double $\cald(S_N)$.

One can show that $A$ is a (braided-)\,commutative symmetric special 
Frobenius algebra.
The irreducible $A$-modules, i.e.\ the ``permutation branes'' in the 
terminology of \cite{schnf}, can be easily determined with the help of theorem
3.2 of \cite{kirI14}. They are labelled by pairs $(\vec\lambda, g)$ where 
$\vec\lambda$ is now an ordered $N$-tuple of labels and $g$ is an element of 
the stabilizer group $\cals_{\vec\lambda}$. One can then apply
the formalism of \cite{fuRs,fuRs4} to compute e.g.\ the boundary states and 
annulus partition functions. The latter turn out to be linear combinations
of twining characters with prefactors that are sums of products of fusion 
rules of the $\calc$-theory. The NIM-rep property  %%integrality
 of the annuli as well as the consistency of the boundary states
is guaranteed by the general theory.

\section{Conclusions}

Besides its computational power and the possibility to obtain general 
proofs of consistency, an additional benefit of our approach is that it
reduces old physical questions to standard problems in algebra and 
representation theory. Here are some interesting examples:
\nxt The classification of CFTs with given chiral data $\calc$ amounts
     to classifying Morita classes of symmetric special Frobenius algebras
     in the category $\calc$. In particular, physical modular invariants of
     automorphism type are classified by the Brauer group of $\calc$.
\nxt The classification of boundary conditions and defect lines is reduced to
     the standard rep\-re\-sen\-tation theoretic problem of classifying modules
     and bi-modules. As a consequence, powerful methods like induced modules
     and reciprocity theorems are at our disposal.
\nxt The problem of deforming CFTs is related to the problem of deforming
     algebras, which is a cohomological question. For the moment, the only
     known results in this direction are rigidity theorems \cite{etno}: a
     rational CFT cannot be deformed within the class of rational CFTs. It will 
     therefore be important to get a better understanding of non-rational CFT.

%%%%%%%%%%%%%%%%%%%%%%%%%%%%%%%%%%%%%%%%%%% 
 \newcommand\wb{\,\linebreak[0]} \def\wB {$\,$\wb}
 \newcommand\Bi[1]    {\bibitem{#1}}
 \newcommand\J[5]   {{\sl #5}, {#1} {#2} ({#3}) {#4} }
 \newcommand\K[6]       {\ {\sl #6}, {#1} {#2} ({#3}) {#4}}
 \newcommand\Prep[2]  {{\sl #2}, preprint {#1}}
 \newcommand\BOOK[4]  {{\em #1\/} ({#2}, {#3} {#4})}

  % N84-LEE-000
 \def\jf    {J.\ Fuchs}

 \def\coma  {Con\-temp.\wb Math.}
 
 \def\cpma  {Com\-pos.\wb Math.}
 \def\fiic  {Fields\wB Institute\wB Commun.}
 \def\ijmp  {Int.\wb J.\wb Mod.\wb Phys.\ A}
 
 \def\josp  {J.\wb Stat.\wb Phys.}

 \def\nupb  {Nucl.\wb Phys.\ B}
 \def\phlb  {Phys.\wb Lett.\ B}
 \def\phrl  {Phys.\wb Rev.\wb Lett.}

 \def\NY     {{New York}}

\end{document}